\documentclass[prc,aps,nofootinbib,superscriptaddress,longbibliography]{revtex4-1}

\usepackage{graphics}

\usepackage{graphicx}

\usepackage{epsfig}

\usepackage{color}
\usepackage{slashed}
\usepackage{multirow}
\usepackage{bm}
\usepackage{footnote}

\usepackage{amssymb, amsmath, latexsym}
\usepackage{eqnarray}

\newcommand{\fm}{\mathrm{fm}}

\newcommand{\asym}{\delta}         
\newcommand{\EPA}{\mathcal{E}}   
\newcommand{\KEA}{\mathcal{T}}   

\begin{document}

\title{Density dependence of the nuclear energy-density functional}

\author{Panagiota Papakonstantinou}
\email{ppapakon@ibs.re.kr}
\affiliation{Rare Isotope Science Project, Institute for Basic Sceince,
Daejeon 34047, Korea}
\author{Tae-Sun Park}
\email{tspark@skku.ac.kr}
\affiliation{Department of Physics, Sungkyunkwan University,
Suwon 16419, Korea}
\author{Yeunhwan Lim}
\email{ylim@tamu.edu}
\affiliation{Cyclotron Institute, Texas A\&M University, College Station, TX 77843, USA}
\author{Chang Ho Hyun}
\email{hch@daegu.ac.kr}
\affiliation{Department of Physics Education,
Daegu University, Gyeongsan 38453, Korea}

\date{\today}

\begin{abstract}
\noindent {\bf Background:} The explicit density dependence in the coupling coefficients entering the non-relativistic nuclear energy-density functional (EDF) is understood to encode effects of three-nucleon forces and dynamical correlations. 
The necessity for the density-dependent coupling coefficients to assume the form of a preferrably small fractional power of the density $\rho$ is empirical and the power is often chosen arbitrarily. Consequently, precision-oriented parameterisations risk overfitting in the regime of saturation and extrapolations in dilute or dense matter may lose predictive power. \\
{\bf Purpose:} 
Beginning with the observation that the Fermi momentum $k_F$, i.e., the cubic root of the density, is a key variable in the description of Fermi systems, we first wish to examine if a power hierarchy in a $k_F$ expansion can be inferred from the properties of homogeneous matter in a domain of densities which is relevant for nuclear structure and neutron stars. For subsequent applications we want to determine a  functional that is of good quality but not overtrained. \\ 
{\bf Method:} For the EDF, we fit systematically polynomial and other functions of $\rho^{1/3}$ to existing microscopic, variational calculations of the energy of symmetric and pure neutron matter (pseudodata) and analyze the behavior of the fits. We select a form and a set of parameters which we found robust and examine the parameters' naturalness and the quality of resulting extrapolations. \\
{\bf Results:} A statistical analysis confirms that low-order terms such as $\rho^{1/3}$ and $\rho^{2/3}$ are the most relevant ones in the nuclear EDF beyond lowest order. 
It also hints at a different power hierarchy for symmetric {\it vs.} pure nutron matter, 
supporting the need for more than one density-dependent terms in non-relativistic EDFs. 
The functional we propose 
easily accommodates known or adopted properties of nuclear matter near saturation. 
More importantly,  
upon extrapolation to dilute or asymmetric matter, it reproduces a range of existing microscopic results, 
to which it has not been fitted. It also predicts a neutron-star mass-radius relation consistent with observations. The coefficients display naturalness. \\
{\bf Prospects:} Having been already determined for homogeneous matter, a functional of the present form can be mapped onto extended Skyrme-type functionals in a straightforward manner, as we outline here, 
for applications to finite nuclei. 
At the same time, the statistical analysis can be extended to higher orders and for different microscopic ({\it ab initio}) calculations with sufficient pseudodata points and for polarized matter. 

\end{abstract}


\maketitle

\section{Introduction}

Among the most successful and widely used models for nuclear structure, are the Skyrme force~\cite{EKR2011} 
and the relativistic mean field (RMF)~\cite{VAL2005} models, 
which provide the basis for a nuclear energy-density functional (EDF) theory. 
They have been applied in the description of many known 
stable and exotic nuclei and the nuclear equation of state (EoS).  
Most traditional RMF and Skyrme force or functional models are fitted to 
properties of experimentally accessible nuclei. 
Consequently, it has been recognized that they are good at reproducing the properties of nuclei in the valley of stability, 
but if one approaches to extremes such as neutron or proton drip lines, or densities much higher or lower than the
saturation density, where precise experimental data are not available, extrapolations must be made with care. 
New experiments on exotic nuclei, as well as astronomical observations, 
help {to} constrain the set of reliable functionals. 
For instance, 
recent observations of $2 M_\odot$ neutron stars~\cite{Dem2010,Ant2013} 
exclude the {nuclear} models giving the maximum mass smaller than 
that value. 

To meet the current challenges, new classes of functionals are being proposed, for example those inspired by the density-matrix expansion~\cite{CDK2008,Bog2011}. 
Extensions and revisions are informed by nuclear-matter constraints~\cite{Dut2012,BSV2008}, new insights from chiral effective field theory (EFT)~\cite{HKW2011,GBD2011,Kai2012} and resummation techniques~\cite{Kai2011,YGL2016}, and by more-stringent stability requirements~\cite{Hel2013}.  
{New optimization procedures and data sets have been employed within the UNDEF project \cite{undef2015, kortel2010, kortel2012, kortel2014}.
}
New, additional {interaction}
terms and forms are introduced for more flexibility and precision~\cite{ADK2006,XPC2016,ZhC2015}. 
The feasibility of describing simultaneously finite nuclei 
and homogeneous {nuclear} matter is continually assessed~\cite{SGS2013,AfA2016}.

Indispensible in all non-relativistic approaches has been the density-dependent term. 
In the original model by Skyrme~\cite{Sky1958} and the seminal work by Vautherin and Brink~\cite{VaB1972} and other early efforts a repulsive three-nucleon contact force was introduced. It counters the attractive two-body interaction and generates saturation. This three-nucleon force is found equivalent to a two-body term whose coupling strength depends linearly on the local density. 
In subsequent parameterizations the linear dependence was replaced with a fractional-power dependence such as $\rho^{1/3}$ or $\rho^{1/6}$ to describe more precisely the compression modulus. 
The Gogny (finite-range) parameterizations are also augmented with such a term. 
The power of the density dependence is considered as a parameter to be fitted or put in ``by hand". 
Nowadays, for more precision, more than one density-dependent terms are considered too, often chosen arbitrarily. Such extensions can produce precise fits, but risk overfitting the parameters, which then may lack predictive power as a consequence. 
So how can we get any guidance regarding the physical values for the density dependence? 

In fact, there exist analytical considerations for interacting Fermi systems which offer clues. 
In the Brueckner theory for homogeneous nuclear matter, when strong interactions between
nucleons are comprised of short-range repulsion and medium-range attraction,
the nuclear potential energy per particle is precisely the sum of powers of the Fermi momentum $k_F$ \cite{FeW1971} (i.e., powers of $\rho^{1/3}$), 
beginning with $k_F^3$.
In a pionless EFT for dilute Fermi systems~\cite{HaF2000}
the energy density of the system up to next-to-next-to-leading order is obtained as a polynomial in terms of $k_F$ also beginning with $k_F^3$. 
Within chiral perturbation theory in the three-loop approximation it has been shown that the saturation regime of nuclear matter is governed by the lowest-order terms, $k_F^3$ and $k_F^4$~\cite{KFW2002}. 
In other words, the cubic root of the density arises naturally in the expression of the energy per particle in the form of $k_F$. 
Not only does it arise naturally, but it cannot be neglected.
We conclude that the fractional-power density dependence, which is indispensible and empirically justified in Skyrme and even Gogny parameterizations but considered dubious in its interpretation~\cite{EKR2010}, in fact arises from a true, though unknown, Hamiltonian within quantum many-body theory and within EFTs. 
One simply considers the functional as a black box, and the generating potential within Hartree-Fock approximation as a pseudopotential, which is close to the spirit of the Hohenberg-Kohn and Kohn-Sham theorems. 

The above observations lead us
to  write the nuclear EDF as a polynomial in $\rho^{1/3}$. 
Such an expansion was explored tentatively in Refs.~\cite{Coc2004a,Coc2004b} with the goal of removing the correlations between the effective mass and the compressibility. 
It was explored again in Ref.~\cite{ADK2006}, where the flexibility of a generalized functional was demonstrated. 
Our point of view is somewhat different. 
We aim explicitly to determine the relevant terms for describing nuclear matter in a wide range of densities relevant for nuclei and neutron stars. 
At this initial stage we are not concerned with spectroscopic precision and therefore we restrict ourselves to homogeneous matter. 
We test to what extent a few low-order terms suffice for realistic results and whether it is possible to establish restricted ranges for the parameters' values already from homogeneous matter, which could be used for precision fits at a later stage. 
Applications of a resulting functional in nuclei are being explored in parallel~\cite{Gil2016,npsm2017} and the related methodology will be outlined in our concluding section.

The manuscript is organised as follows. In Sec.~{\ref{Sec:background} we elaborate on the reasoning that leads to the present functional form. 
In Sec.~\ref{Sec:theory} we present the form of the proposed functional and explain the fitting procedure. 
In Sec.~\ref{Sec:results} we present our results and demonstrate a power hierarchy which justifies the power expansion to low order, {especially for symmetric matter}, and the coefficients' naturalness. 
As a first application, we solve the Tolman-Oppenheimer-Volkoff (TOV) equations and obtain the mass-radius relation of neutron stars. 
A summary and perspectives are given in Sec.~\ref{Sec:summary}. 

\section {Why powers of $k_F$\label{Sec:background}} 

Some of the observations in this Section, especially as regards EFT, will rightly appear speculative.  
Let us therefore make our point clear from the beginning: Taking indications from effective theories (as elaborated below), in this work {\em we assume} that we can write the potential energy per particle as a low-order expansion in $k_F$. 
It would be very interesting to establish such an expansion, as it would eliminate the uncertainty regarding the most important powers in the density dependence of the EDF. 
If in our subsequent analysis (Sec.~\ref{Sec:results}) we find no numerical evidence for such a low-order expansion (for example, if the $\rho$ term gives interchangeable fits with, say $\rho^{1/3}$), our assumption is of course rendered moot. 
However, it turns out this is not the case. Especially for symmetric nuclear matter, it turns out that the expansion might converge fast. 
Let us therefore present the physical reasoning which leads us to expect that a low-order expansion might work for nuclear matter within the regime of interest, i.e., within an order of magnitude below or above the saturation density  $\varrho_0 \approx 0.16~$fm$^{-3}$.
We certainly hope that the interested reader will find inspiration to explore and refine the present ideas in more depth and will find our numerical results useful in such a pursuit. 

Following a textbook example on nuclear matter~\cite{FeW1971}, let us assume a local interaction between nucleons with repulsive hard core and a longer-range attractive part (of finite depth $V_0$) of ranges $r_c$ and $r_a$, respectively. 
The contribution of the repulsive core to the potential energy per particle is given by Brueckner theory as a sum of terms proportional to $(k_Fr_c)^3$, $(k_Fr_c)^4$, etc.~\cite{FeW1971}, 
and converging slowly for $k_Fr_c\approx1$. The contribution of the attractive part to the energy per particle is given in closed analytical form involving trigonometric functions and integrals of $k_F r_c$ and $k_Fr_a$~\cite{FeW1971}, which can also be expanded in ascending powers of $k_F$,
$k_F$, the lowest power surviving is, in 
\begin{equation} 
\EPA_a = -\frac{V_0k_F^3}{2\pi} (r_a^3-r_c^3) - \frac{9V_0}{2\pi} \sum_{m=2}^{\infty} 
\frac{(-1)^{m+1}}{s_m} \left[  (k_Fr_a)^{2m+1} - (k_Fr_c)^{2m+1} \right] 
\label{Eq:Bth} 
,
\end{equation} 
where the denominator 
\[ s_m=(2m-1)! (2m+4)(2m+2)(2m+1)^2 / 2^{2m+1} \] 
diminishes quickly with increasing $m$. In this picture, the even powers of $k_F$ arise from the repulsive part only. 

The importance of the $k_F^3$ and $k_F^4$ terms for obtaining the empirical saturation regime of symmatric matter was shown explicitly in Ref.~\cite{KFW2002} within the three-loop approximation of chiral perturbation theory. 
In the very particular case of extremely dilute Fermi systems, the expression for the energy per particle has been obtained, e.g., in Ref.~\cite{HaF2000} as a polynomial expansion in $k_F$, where the expansion coefficients depend on the scattering lengths and the effective ranges, plus logarithmic functions, arising from three-fermion forces. 
That is in fact an analytical form we will explore in this work, but with the expansion coefficients treated as free parameters. 
(In the end we conclude from our fits that the inclusion of a logarithmic term is not a necessity.) 
At this point we must expose our reasoning for accepting a dilute regime as a starting point for our investigation.  

Notwithstanding the preceding arguments for a polynomial expansion, saturated matter is arguably not at all dilute: The effective range of the interactions is of the order of the interparticle distance, while the bare scattering length is much longer. 
On the other hand, arguments can be made for considering near-saturated matter dilute with respect to certain physics of relevance. Such would be the case within an effective theory without pions but only heavier mesons. 
Since pion is a pseudo-scalar meson, its mean field does not appear in nuclear matter
unless the matter density is high enough to allow pion condensations~\cite{Mig1978}.
In addition, the expectation value of the one-pion-exchange potential vanishes in nuclear matter.
Thus pionic contributions to the energy density are through
loops and multi-pion exchanges, and one may postulate that their average effect
is a modification of the couplings and masses among nucleons and heavy mesons.
Since the Fermi momenta in the measurable nuclear systems and even in neutron stars are smaller than the next heavy-meson mass, namely $m_{\rho}$ (approximately $775$~MeV, or 4~fm$^{-1}$),
one may treat $m_{\rho}$ as a large scale and envision an effective Lagrangian in powers of $k_F/m_\rho$. 
Of course, neglecting pions, a precise matching with nature at threshold region is neither possible 
nor meaningful.
Instead, one would have to fit the Lagrangian coefficients to 
data 
and confirm the accuracy of the approach for describing dense matter  {\em a posteriori}. 
Our approach originates in this idea. For this reason we will examine the naturalness of our fitted coefficients with respect to a $k_F/m_{\rho}$ expansion. 
Let us add that in the RMF models nuclear saturation is obtained from the balance between 
the attractive force by the exchange
of sigma mesons and the repulsive force by omega mesons, while 
pions are not explicitly included. 
{$\rho$ meson is added to reproduce (or control) 
the asymmetric nuclear matter properties better than the conventional
$\sigma$ and $\omega$ RMF models.} 
The success of RMF as well as Skyrme models may imply that the major properties of dense nuclear matter are controlled by short-range forces.

We now comment on the density-matrix expansion (DME)~\cite{NeV1972}, which is popular in recent optimizations of the nuclear EDF. The DME skips some low-order powers of $\rho^{1/3}$.  
This could be because it considers only statistical correlations in the expression for the two-body density matrix, namely the exchange term determined by the off-diagonal one-body density matrix, but it neglects an irreducible two-body dynamical correlation.
The correlation function vanishes when the wave function is a single Slater determinant (free Fermi gas) but constitutes a significant correction in the presence of short-range correlations, which can be treated within a variety of quantum many-body methods~\cite{GR80}. We are not actually proving here that the correlation function will generate the missing terms of $\rho^{1/3}$, but our observation that such terms do arise in EFT and Eq. (\ref{Eq:Bth}) may motivate 
further investigations.

\section{Methodology\label{Sec:theory}}

\subsection{Form of the energy density functional} 

The present Ansatz for 
the energy per particle, 
except the Coulomb energy, in the case of a homogeneous system of 
nucleons with proton density $\rho_p$ and neutron density $\rho_n$, reads 
\begin{equation} 
\begin{aligned}
\EPA (\rho,\asym ) =  \frac{E(\rho,\asym )}{A} = \KEA (\rho ,\asym ) 
                       + \sum_{i=0}^{3} c_i(\asym) \rho^{1+i/3} 
                      + c_{\ln}(\asym) \rho^2 \ln [\rho~{\fm^3}] 
\label{Eq:FormC}
\end{aligned} 
\end{equation} 
where 
we have introduced the total density $\rho = \rho_n+\rho_p$ and the asymmetry $\asym = (\rho_n - \rho_p ) /\rho $. 
The 
kinetic energy per particle is given by 
\begin{equation} 
\begin{aligned}
& \KEA = \KEA_p + \KEA_n \,,
 \\
& \KEA_{p,n} = \frac{3}{5}\frac{\hbar^2}{2m_{p,n}} x_{p,n}^{5/3}(3\pi^2\rho )^{2/3} 
\end{aligned}
\end{equation} 
with $x_{p,n} \equiv \rho_{p,n}/\rho$. We motivated such a form in Sec.~\ref{Sec:background}. 
We are interested in determining the most relevant terms in this expansion and whether any hierarchy can be inferred. 
At present we examine up to the $i=3$ term, but in general higher-order powers can be considered and explored as well. 

We may rewrite Eq.~(\ref{Eq:FormC}) as  
\begin{equation} 
\EPA = \KEA + \sum_{i=0}^3 \EPA_i + \EPA_{\ln}  
\label{Eq:FormE} 
\end{equation} 
where the dependence on the density $\rho$ and the asymmetry $\delta$ of various terms
should be understood.
The $\EPA_3$ and $\EPA_{\ln}$ terms are in effect a single term 
$
c_3' \rho^2\ln  [\rho / \rho_x]
$, 
if we define an unknown reference density value $\rho_x$. 
For purely practical reasons and without loss of generality we prefer to work with two separate terms.

We proceed to specify the asymmetry dependence of the potential energy, $\EPA-\KEA$. 
We stress that the dependence we adopt does not affect our fits at all. 
It only enters our final comparison with the results of chiral EFT in asymmetric matter, which have large error bands, and the modelling of the neutron-star mass-radius relation, which is not precisely determined either. 
{For our comparisons we therefore assume the standard quadratic dependence, which is generally adopted, 
for example within the generalized liquid drop model \cite{ducoin2010,ducoin2011} and 
in recent analytical parameterizations of the chiral EFT results \cite{DSS2014, DHS2015}.
}
Then we can write 
\begin{equation} 
c_k(\asym ) = \alpha_k + \asym^2\beta_k   \quad ; \quad k=i \,\,\mathrm{or}\,\, k=\ln. 
\label{Eq:CandA}
\end{equation}  
There certainly exist open issues regarding the asymmetry dependence of the nuclear EoS~\cite{Kai2012}, but they lie beyond the scope of the present manuscript. 

The functional can be rewritten in the form of a Skyrme functional, with the terms assigned as in Table~\ref{tableSk}, which shows explicitly the powers of the Fermi momentum corresponding to each term.    
\begin{table} 
\begin{center}
\begin{tabular}{|c|c|c|}
\hline
$\KEA$ & kinetic en. & $k_F^2$ \\ 
$\EPA_0$ &  $t_0$ & $k_F^3$ \\ 
$\EPA_1$ &  $t_3$, $a=1/3$ & $k_F^4$ \\ 
$\EPA_2$ &  $t_1, t_2$ ; $t_3'$, $a'=2/3$ & $k_F^5$ \\ 
$\EPA_3$ &  $t_3''$, $a''=1$ & $k_F^6$ \\ 
$\EPA_{\ln}$ & special  & $k_F^6\ln k_F$ \\
\hline 
\end{tabular} 
\end{center}
\caption{\label{tableSk}Correspondence of 
the terms in Eq.~(\ref{Eq:FormE}) to conventional 
Skyrme-functional terms and to powers of Fermi momentum.} 
\end{table} 
We note the presence of more than one density-dependent terms in such a 
corresponding Skyrme functional: a fractional-power one ($\EPA_1$), a linear one ($\EPA_3$), 
 and optionally a second fractional-power one ({contributing to} $\EPA_2$) and a linear-logarithmic one ($\EPA_{\ln}$), which should be considered together with the linear one. 
The presence of more than one density-dependent 
terms renders this form more flexible than the traditional Skyrme functionals. 
Indeed, generalized Skyrme functionals with more than one density-dependent couplings have been explored in Refs.~\cite{ADK2006,ZhC2015}. 
The differences in our case are that the terms are not arbitrarily chosen and that we determine the parameters in homogeneous matter before applications to nuclei. 
We note finally that the present functional accommodates the empirical fractional-power density dependence of the nuclear symmetry energy, 
with power $a = 0.72 \pm 0.19$ \cite{russotto2016}.

If our tentative effective-theoretical arguments in Sec.~\ref{Sec:background} have any merit, the expansion coefficients should display naturalness with respect to the expansion variable $k_F/m_{\rho}$. 
Noting that, at zero temperature, $\rho = \frac{\nu}{6\pi^2}k_F^3$, where $\nu =4$ for symmetric nuclear matter (SNM) and $\nu =2$ for pure neutron matter (PNM), we write
\begin{equation} 
\EPA_i(\rho, \delta ) = c_i (\delta ) \rho^{1+i/3} = \left[ \left( \frac{\nu}{6\pi^2} \right)^{1+i/3} 
 c_i(\delta )  m_{\rho}^{2+i} \right] 
m_{\rho} \left( \frac{k_F}{m_{\rho}} \right)^{3+i}   .  
\end{equation}  
We note that $k_F/m_\rho \simeq 1/3$ for saturated SNM.
We will examine whether the dimensionless parameters 
\begin{equation} 
c_i^{dim}(\delta ) = \left( \frac{\nu}{6\pi^2} \right)^{1+i/3} c_i(\delta )  m_{\rho}^{2+i}
\label{Eq:cdim} 
\end{equation} 
are of the same order of magnitude. 

\subsection{Fitting method\label{Sec:Fit}} 

Having defined the form of the functional, we proceed to determine and analyze the unknown parameters. 
For nuclear-structure applications, one may follow the usual procedure of 
fitting to nuclear properties as well as saturation properties of nuclear matter. 
As already elaborated, our objective is different: we are interested in validating and analyzing our Ansatz in homogeneous matter first.  
{Also important is to not fit all five parameters blindly, but examine which are the most important ones, whose values do not depend strongly on the fitting procedure and may retain some physical content. 
We thus inspect the fits of all possible combinations of $1-5$ parameters.}

The most appropriate set of 
pseudodata {for our purposes} would include both symmetric and asymmetric matter. Therefore we use 
the Akmal-Pandharipande-Ravenhall variational results, which are based on the Argonne V18 and Urbana potentials and available for both SNM and PNM~\cite{APR1998}. 
This set of
 pseudodata will be denoted as APR. 
{For the purpose of confirming our statistical fit analysis, the fitting has been repeated with the Friedman-Pandharipande (FP) pseudodata set~\cite{FrP1981}, within the same density domain as the fit to the APR set.}  
The FP calculations were based on the Argonne V14 potential. 
The APR data are considered an improvement over the FP data, because APR took into account the most accurate two- and three-nucleon interaction until those days, relativistic boost interaction, and the phase transition in the high density region. Therefore they constitute our main set. 

We fit the functional form to the SNM data 
to obtain the parameters $c_i(0)$ and then to the PNM data to obtain $c_i(1)$. 
We will test our results against the chiral EFT results of Ref.~\cite{DSS2014}, which we will denote as DSS, 
available for $\asym=1,0.9,0.8,0.7$. 
The necessary interpolation to asymmetric matter is possible via Eq.~(\ref{Eq:CandA}). 
No fits are performed to the DSS data. 
We will also compare with the SLy4 Skyrme functional, which was partly constrained by microscopic results for homogeneous matter~\cite{Cha1998}.

For the fits to the APR (or FP) set, and separately for SNM and PNM, we proceed as follows:  
We make use of all available pseudodata points $(\rho_j,D_j)$ for 
a given asymmetry value $\delta =0,1$. We perform a least-squares fit by minimizing 
\begin{equation} 
\chi^2 (\delta )= \sum_{j} \exp\{-\beta \rho_j / \varrho_0 \} 
\left(  \frac{\EPA (\rho_j) - D_j}{\KEA (\rho_j)} \right)^2 \, ; 
\quad \beta \geq 0 
\label{Eq:chidef} 
\end{equation} 
with $\varrho_0$=0.16~fm$^{-3}$. 
A dependence of the data-points set $\{ j\}$ 
and related values on $\delta $ is implied. 
For the fits we use the multiparameter regression routine of the 
GNU Scientific Library~\cite{GSLurl}. 
{We next proceed to explain the above choice for the cost function.}

The division of the cost function
in Eq. (\ref{Eq:chidef}) 
with the kinetic energy allows us to {increase the weight of the comparatively small and disfavored contributions of $\EPA (\rho_j) - D_j$ at lower densities, without introducing arbitrary weight functions. 
Some further weighting} is necessary nonetheless, owing to the nature of the data: 
As pointed out in Ref.~\cite{APR1998}, the pseudodata show a discontinuity 
at some value of density near 0.2 or 0.3 fm$^{-3}$. The authors recommend 
and use a different parameterisation for the ``low-density phase" (LDP) 
and the ``high-density phase" (HDP). 
A phase transition at similar density 
is discussed in \cite{PKL2016}. 
Indeed, we have found that an unweighted fit 
to the pseudodata provides an overal good description of the data, 
but does not reproduce as precisely as desired the saturation point (cf. Table~\ref{Table:CI}). 
On the other hand, neglecting the HDP altogether would not be sufficient 
for constraining the higher-order terms of the expansion (there are few low-density data) 
and is not recommended for an application to neutron stars~(cf. Fig.~\ref{Fig:MR}).  
In order to examine specifically the saturation region and the dilute-density regime, we can further adjust 
the weight put on it via the parameter $\beta$ introduced 
in Eq.~(\ref{Eq:chidef}). 
For meaningful comparisons between results with different values of $\beta$ 
it is then pertinent to introduce the normalized quantity 
\begin{equation}
\chi^2_n (\delta ) = \chi^2 (\delta ) \left[ \sum_{j} \exp\{-\beta \rho_j / \varrho_0 \} \right]^{-1} 
\label{Eq:chinorm} 
\end{equation} 
to remove a trivial decrease of $\chi^2$ as $\beta $ is increased. 

{Elements of information theory are utilized to corroborate the stiffness or sloppiness of the model parameters, see, e.g., Ref.~\cite{NiV2016} for a recent application in nuclear density functional theory. 
We first define the Hessian matrix of the model as
\begin{equation} 
H_{kl} (\delta ) = \frac{1}{2N_c}\frac{\partial^2}{\partial\log{\tilde{c}_k}\partial\log{\tilde{c}_l}} 
\int_{\varrho_0 /5}^{5\varrho_0} d\rho   \left[ \frac{\EPA(\rho ,\delta ; \tilde{c}_k) - \EPA(\rho ,\delta ; \tilde{c}_k^{(f)})}{\EPA (\varrho_0,0)}   \right]^2
\label{Eq:Hess},  
\end{equation} 
where $N_c$ is the number of fitted parameters. 
The dimensionless parameters $\tilde{c}_k$ are defined via (cf. Eqs.~(\ref{Eq:FormC}), (\ref{Eq:FormE})) 
$\tilde{c}_i(\delta ) = c_i (\delta ) \varrho_0^{1+i/3}/\EPA (\varrho_0,0), \,
\tilde{c}_{\ln}(\delta ) = c_{\ln} (\delta ) \varrho_0^{2}/\EPA (\varrho_0,0)$.
The integral defining the Hessian quantifies the change in model behaviour, within the representative range of densities $(\varrho_0/5,5\varrho_0)$, as parameter values $\tilde{c}_k$ vary around the fit results $\tilde{c}_k^{(f)}$. The presence of Hessian eigenvalues spread over many orders of magnitude will signify the presence of sloppy parameters and hence the possibility to eliminate at least one combination of those without much spoiling the model quality. 
That's quite relevant for our purposes: It will signal that additional parameters are superfluous. If, on the other hand, the eigenvalues are of comparable magnitudes, then the model is stiff and might miss relevant physics. 
We will therefore look at the behaviour of the eigenvalues as we add more parameters to the fits.}

\subsection{{Application in neutron stars}} 
As a first application we will consider the mass and radius of neutrons stars. 
For this purpose we will solve the TOV equations, 
which for given central density read 
\begin{equation}
\begin{aligned}
\frac{dp}{dr} &= - \frac{G(M(r)+4\pi r^{3} p/c^{2})(\varepsilon + p)}
{r(r-2GM(r)/c^{2})c^{2}}, \\
\frac{dM}{dr} &= 4\pi \frac{\varepsilon}{c^{2}}r^{2},
\label{eq:tov}
\end{aligned}
\end{equation}
where $r$ is the radial distance from the center,
$M(r)$ is the enclosed mass of a neutron star within $r$, and
 $p$ and $\varepsilon$ represent pressure and energy density respectively.

{We will solve the above equations for the energy-density parameterizations obtained from our fits. 
Acceptable functionals should allow for neutron star masses to reach the value of two solar masses~\cite{Dem2010,Ant2013} and produce mass-radius relations within the currently accepted constraints deduced from X-ray burst data~\cite{SLB2010}.}

\section{Fitting analysis, results and application\label{Sec:results}} 

\subsection{Fitting analysis} 

Fits have been performed for the 31 possible combinations of one, two, ..., 
or five non-zero constants from the set $\{ c_k(\delta )\}$ ($\delta =0,1$). 
Table~\ref{Table:Chi} lists the values of the normalized cost 
function $\chi^2_n$ for a few selected combinations of fitted parameters and 
$\beta=0,1/2,1$. 
For completeness, 
results for all examined combinations of parameters and additionally for $\beta=3/2$ are 
provided {in the Appendix, Table~\ref{Table:Chi2}}.

First we discuss the successive inclusion of non-zero parameters, shown in the 
first row of each block of results, namely data rows 1,3,9,12. 
It is of course a trivial result that, as we include more 
and more parameters, the fits get better. 
{However, a saturation of the fit quality is observed in the case of SNM when 3 parameters are included (stable $\chi^2$), as long as the lowest-order term $c_0$ is included.  
} 

A hierarchy of terms, where the lower-order ones are more important than the higher-order ones, 
is inferred from the present results:
\begin{itemize} 
\item  
Generally speaking, for a given number of parameters, the sets which include the $k=0$ term give better fits than those which do not. There are a few exceptions and mostly for low $\beta$. 
\item 
In the majority of cases, if we replace the $k=1$ term with the $k=3$ term we get noticeably higher $\chi_n^2$ values. 
This result is in concordance with the preference for Skyrme functionals with a fractional-power, rather than linear, density dependence.  
\item 
If we use only two parameters, the sets of two low-order parameters produce better fits than the sets of two higher-order parameters. 
For example for $\beta=1$ one may arrange the sets 
from {the} best to worst as follows: 
For SNM, $k=(0,1)$, $(0,2)$, $(0,3)$, $(1,2)$, $(1,3)$,
 $(2,3)$, $(3,$ln).\footnote{We do not discuss sets which include the $k=$ln term without the $k=3$ term, because they imply an arbitrary reference density $\rho_x=1$fm$^{-3}$.} 
For PNM the order is the same except that $k=(0,2)$ is better than $(0,1)$. 
\item 
In fact for smaller $\beta$ the $k=3$ term in PNM seems more efficient. 
The inclusion of a linear dependence in Skyrme functionals might be recommended especially for dense-matter applications. 
We note that the discontinuity of the data may contaminate the systematics of the low-$\beta$ fits. 
\item 
For three parameters, we found that the smallest $\chi_n^2$ are {generally} obtained without the logarithmic term. 
\end{itemize}

\begin{table*}[h]
\begin{center}
\begin{tabular}{|r|c|c|c|}
\hline 
    &   $\beta=0$ & $\beta=\frac{1}{2}$ & $\beta =1$ \\ 
\hline\hline 
    &     SNM   PNM   &   SNM   PNM   &  SNM   PNM   \\ 
\hline
 $k=0$ & 1.595335  0.397036 & 0.930742  0.171609 & 0.490650  0.071632 \\
 $k=1$ & 1.801776  0.346198 & 1.527834  0.223333 & 1.089477  0.138133 \\
\hline
 $k=0,1$ & 0.013044  0.022028 & 0.003866  0.007482 & 0.001151  0.001566 \\
 $k=0,2$ & 0.009356  0.005804 & 0.012267  0.001864 & 0.009435  0.000719 \\
 $k=0,3$ & 0.041156  0.002160 & 0.047771  0.003059 & 0.035831  0.003220 \\
  $k=1,2$ & 0.085297  0.005936 & 0.108696  0.009991 & 0.090303  0.010973 \\
 $k=1,3$ & 0.175982  0.014031 & 0.216418  0.022334 & 0.183405  0.023312 \\
 $k=2,3$ & 0.342376  0.031821 & 0.440564  0.048252 & 0.398009  0.050970 \\
\hline
 $k=0,1,2$ & 0.005009  0.003287 & 0.002588  0.001781 & 0.001016  0.000529 \\
 $k=0,2,3$ & 0.006453  0.002055 & 0.004070  0.001540 & 0.001284  0.000636 \\
 $k=1,2,3$ & 0.021528  0.005183 & 0.018591  0.005162 & 0.008571  0.003018 \\
\hline
 $k=0,1,2,3$ & 0.001616  0.000163 & 0.001731  0.000188 & 0.001015  0.000138 \\
 $k=0,1,2,(\frac{7}{3})$ & 0.001420  0.000115 & 0.001597  0.000136 & 0.001016  0.000112 \\
 $k=0,1,2,(\frac{8}{3})$ & 0.001268  0.000098 & 0.001472  0.000106 & 0.001009  0.000092 \\
 $k=0,1,2,$ln & 0.001314  0.000094 & 0.001510  0.000107 & 0.001011  0.000092 \\
 $k=0,1,2,(\frac{1}{6})$ & 0.002277  0.000462 & 0.002072  0.000415 & 0.000977  0.000221 \\
\hline \hline 
\end{tabular} 
\end{center}
\caption{\label{Table:Chi}$\chi^2_n$ values for the indicated fits (selections of $\beta$ and non-zero $c_k$) of expression (\ref{Eq:FormC}) to the APR pseudodata in SNM and PNM. 
An extended version of this Table is given in the Appendix, Table~\ref{Table:Chi2}. 
}
\end{table*} 

Let us evaluate further the necessity for higher-order terms. 
For this we consider fitting the $k=0,1,2$ terms along with one more term whose form may be $\rho^2$ ($k=3$), $\rho^2\ln [\rho $~fm$^3]$ (logarithmic term), 
$\rho^{7/3}$ (next-order term to $k=3$), $\rho^{8/3}$ (approximate symmetry-energy dependence within Dirac-Brueckner-Hartree-Fock~\cite{Sam2012}) or $\rho^{1/6}$ (popular in Skyrme functionals).  
(A systematic inclusion and examination of higher-order terms, such as $\rho^{7/3}$, $\rho^{8/3}$, is deferred to future work.)  
The results in the last rows (last block) of Table~\ref{Table:Chi} demonstrate that the quality of the fit is 
almost unaffected by the choice of 4th term. 
An interesting exception is that the popular $\rho^{1/6}$ term generally gives a worse fit. 
The most precise fit is of course provided by 5 terms. 
However, the resulting values for the coefficients $c_i$ are found radically different from those obtained with 4 parameters or fewer. 
{From the above we can {infer} that two {(SNM)} or three {(PNM)} terms are essential, that the role of a fourth high-order term is simply to refine the fits, and last but not least, that a fifth term cannot be constrained by the pseudodata, i.e., it may lead to overfitting, which is not desired. 

The above conclusions are supported by an analysis of the Hessian spectrum, which we now discuss briefly, for both the APR and the FP data sets. 
The Hessian matrix of Eq.(\ref{Eq:Hess}) has been evaluated and its eigenvalues computed for the various fits. 
Figure~\ref{Fig:Hess} 
shows, for each fit, the ratio of the highest (in absolute value) eigenvalue of $H$ to the lowest one (in absolute value). 
Each point corresponds to a specific combination of 2, 3, 4, or 5 terms included in the fit. For example, in the case of 3 parameters (middle panels), there are, for each value of $\beta$, $_{5}C_{3} -3= 7 $ points for PNM and 7 for SNM (we exclude the three combinations where $c_{\ln}$ appears without $c_3$).  
Figure~\ref{Fig:Hess} indicates the combinations giving the lowest and highest ratios. 
For example, the label $k=(0,1,3)$ means that the parameters $c_0,c_1,c_3$ were fitted to obtain the adjacent points. If the label is centered (as, e.g., in Fig.~\ref{Fig:Hess}(b) and (e)~), the label refers to the points on either side of the label.  

\begin{figure*} 
\begin{center}
\includegraphics[scale=0.7]{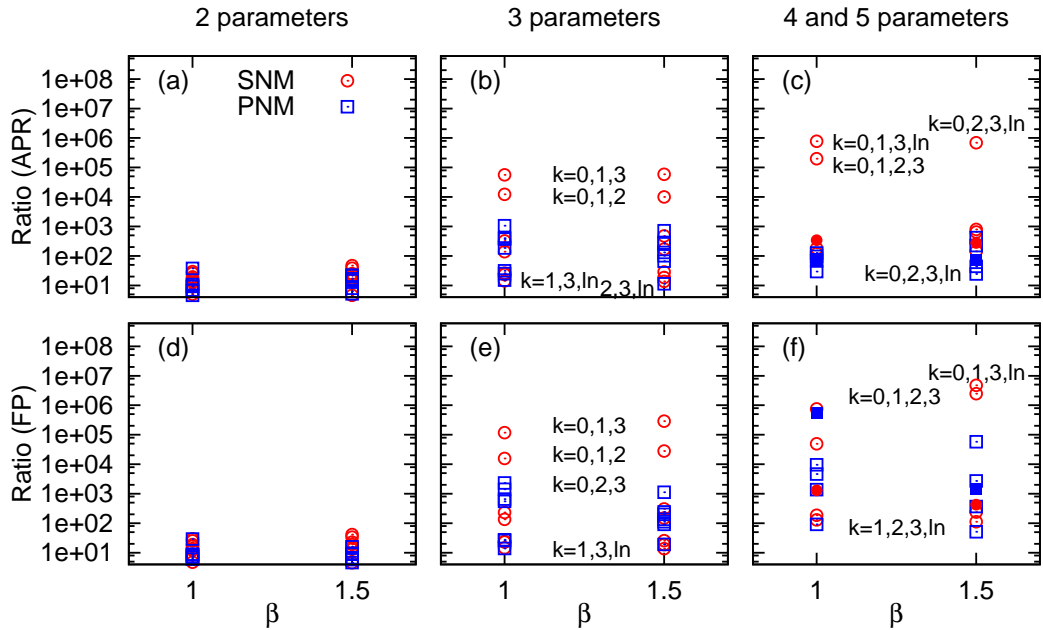}\nolinebreak[4]\hspace{-60mm}\\[-110mm] 
\caption{\label{Fig:Hess} 
{Ratio of the highest to the lowest Hessian eigenvalues (in absolute value) for various fits with 2,3,4, or 5 parameters. Full symbols correspond to 5 parameters. 
(a), (b), (c): Fits to APR data. 
(d), (e), (f): Fits to FP data. 
For each dataset, asymmetry and $\beta$ value, a point corresponds to a specific combination of parameters used. 
In the cases of 3 or 4 parameters, the combinations giving the highest and lowest Hessian ratios are indicated.}
}
\end{center}
\end{figure*} 

For a given number of terms, $N_c$, the highest ratios are provided by fits with low-order parameters, meaning that an optimal choice of $N_c-1$ low-order parameters is possible. 
The lowest ratios correspond to fits where all terms are very important, either fortuitously, or because the parameters are not optimally chosen. 
The results reveal that two parameters are barely sufficient to retain flexibility. 
The same holds for three or even four {\em high-order} parameters}, i.e., when low-order terms ($k=0$ or $1$) are omitted. Therefore the low-order terms appear more physical.   

Again we observe the different behaviour of SNM and PNM. For SNM 3 parameters 
seem {to be} sufficient, but for PNM the parameters are generally stiffer. 
It will be interesting to examine the PNM thoroughly with more terms and statistical analyses, and on richer sets of pseudodata. 
This shall be the subject of future work. 
Because of the above {observations}, and to avoid overfitting, a fifth term (high-order or logarithmic) is omitted in what follows.  
For our initial applications the present fits are already of good quality. 

In Fig.~\ref{Fig:EoAa}, representative results of fits for the EoS are shown, along with the APR pseudodata, the DSS results from chiral EFT, and the SLy4 functional. The almost-linear trend with respect to $\rho^{1/3}$, especially for SNM, is evident. The effect of including higher-order terms appears indeed minimal in SNM. 
In the case of PNM, the fit with four parameters not only reproduces well the pseudodata (APR), but also the results of chiral EFT (DSS), to which it has not been fitted, and at low density, where no pseudodata exist.  
This result is not trivial~\cite{YGL2016}, as the comparison with traditional functionals shows, for example SLy4 in Fig.~\ref{Fig:EoAa}. 
\begin{figure*} 
\begin{center}
\includegraphics[scale=0.51]{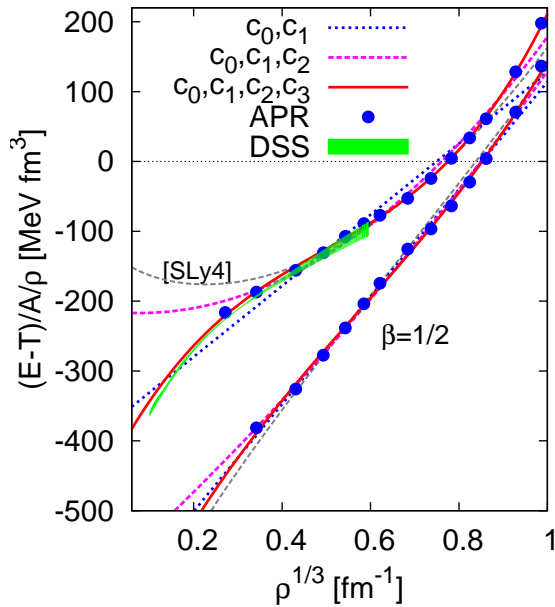} 
\caption{\label{Fig:EoAa} 
Representative results ($\beta = 1/2$) of fits with two, three, or four parameters, as indicated, along with the APR pseudodata, the DSS results from chiral EFT and the SLy4 functional. 
Shown is the potential energy per particle divided by the density as a function of $\rho^{1/3}$, in SNM and PNM. 
%
}
\end{center}
\end{figure*}

\subsection{{Resulting functionals, naturalness, and application}} 

{Having verified the relevance and quality of the low-order fits and having set the number of desired parameters to four, we now proceed to determine specific parameterizations, to use as starting points to applications.} 

In what follows we choose as our main set of results 
the 4-parameter low-order form, with $k=0,1,2,3$.
Table \ref{Table:CI} shows the resulting values of $c_k(0)$ and $c_k(1)$ for $\beta =0,0.5,1$. 
The values obtained for the properties of nuclear matter at saturation density are also shown.
For the fitted parameters, the saturation properties are reasonable, but not precisely equal to the known or adopted values, which is not surprising: 
The pseudodata near saturation are few and the set includes a kink at higher densities.  
However, it is straightforward to make adjustments to the SNM parameters so as to obtain any desired set of values of SNM properties. 
In fact, one can do away with the pseudodata of SNM and adjust the 
parameters $c_i(0)$ to chosen SNM properties by solving simple algebraic equations. If this procedure produces similar parameters as the fitting, for the low-order terms, the present expansion Ansatz will be validated further.  
For the purpose of demonstration, we presently explore two options, as follows. 

The first option is to adjust the SNM parameters to a saturation density $\varrho_0=0.16$~fm$^{-3}$, binding energy per nucleon  
$\EPA_0 =-16.0$~MeV, incompressibility $K_{\infty}=240$~MeV, and nucleon effective mass $m^{\ast}/m=0.7$. 
The effective mass in this case is calculated by assuming that the $c_2$ term is entirely of the form $\rho\KEA$, which is of course a non-binding and arbitrary choice for the sole purpose of the present demonstration: 
The portion of $c_2$ term coming from non-local terms ($t_1,t_2$) cannot be determined from data on unpolarized homogeneous matter}.\footnote{{Explorations in finite nuclei are in progress~\cite{Gil2016,npsm2017}.}} 
The resulting coefficients for SNM are shown in the two rows labeled "ad-1" of Table~\ref{Table:CI}. 
They do not deviate much from the fitted values.  
In fact, their similarity to the values obtained with $\beta =1$ 
shows that the variable $\beta$ properly puts {more} weight on the lower-density data
{than the high-density regime}. 
For the PNM we presently chose the same coefficients as the fit with $\beta =1/2$, so as to retain some weight on the high-density regime. 

The second option is to simply set $c_3(0)=0$ and determine the other three parameters from the above values for $\varrho_0, \EPA_0, K_{\infty}$. 
This time we make no assumption for the effective mass, which remains unconstrained. 
Note that $c_3(0)$, being a sloppy parameter, changes sign when $\beta$ is varied. The value of $c_3(0) =-0.00$ (to that precision) is in fact obtained for a fit with the acceptable weight function $\beta=0.97273$. 
The resulting parameters, as well as the PNM parameters corresponding to $\beta = 0.97273$, are also listed in Table~\ref{Table:CI}, labeled ``ad-2". The agreement of $c_i$ with $i=0,\, 1$ for all the last three sets of parameters corroborates the robustness of our approach. 

{We should stress that, for all sets, the $J$ and $L$ values are not input values, but obtained. 
The values are within the currently proposed constraints \cite{Dut2012,LL_apj2013}.} 

\begin{table*}[t] 
\begin{center}
\begin{tabular}{|c|c| r r r r | rrr|}
\hline 
%
$\beta$&Matter  &$c_0$~~&$c_1$~~&$c_2~~$& $c_3~~$ 
 &  $\varrho_0$ & $\EPA_0$ & $K_{\infty}$ 
 \\ & & & & & & & $J$ & $L$  \\  
\hline 
\multirow{2}{*}{$0$}  
  &SNM  &  $-863.36$&   $1945.05$&   $ -2060.20$&   $1129.96$  & 0.178 & $-15.4$ & 215 \\ 
  &PNM  &  $-483.96$&   $1433.54$&   $ -2119.68$&   $1385.22$  & & 34.2 & 55.9   \\  
\hline
\multirow{2}{*}{$\frac{1}{2}$} 
  &SNM  &   $-753.98$&$1389.20$&   $ -1171.03$&   $678.87$       & 0.177 & $-15.8$ & 234 \\
  &PNM  &   $-451.91$&   $1254.32$&   $ -1812.62$&   $1221.33$  & & 34.4 & 56.0    \\
\hline
\multirow{2}{*}{$1$} 
  &SNM  &  $-613.13$&   $620.22$&$ 154.72$&   $-46.05$             & 0.171 & $-16.1$ & 247 \\ 
  &PNM  &  $-408.56$&   $991.76$&  $-1323.81$&   $937.96$ & & 34.0 &  54.9  \\
\hline\hline 
\multirow{2}{*}{ad-1} 
  &SNM  &  $-648.72$&$676.25$&  $ 200.92$&  $-98.73$              & 0.160 & $-16.0$ & 240 \\
  &PNM  &  $-451.91$&  $1254.32$&  $-1812.62$&  $1221.33$ & & 32.8 & 47.9     \\
\hline\hline 
\multirow{2}{*}{ad-2} 
  &SNM  &  $ -664.52 $&$  763.55 $&  $   40.13 $&  $   0.00$              & 0.160 & $-16.0$ & 240 \\
  &PNM  &  $-411.13$&  $ 1007.78$&  $ -1354.64$&  $ 956.47$ & & 33.5 & 50.5      \\
\hline
\end{tabular} 
\end{center}
\caption{\label{Table:CI}Coupling constants in SNM ($c_k(0)$) and PNM ($c_k(1)$) in units MeV~fm$^{3+k}$ obtained from fits with the indicated $\beta$ values to the APR pseudodata and corresponding bulk-matter properties: saturation density $\varrho_0$ in fm$^{-3}$;  
energy per particle at saturation $\EPA_0$, 
incompressibility $K_{\infty}$, symmetry energy $J$ and slope parameter $L$ in MeV. 
The values for the functional which was adapted to SNM saturation properties are shown in the last block (``ad", see text).
{For all the sets, $J$ and $L$ are not input values but obtained results.}}
\end{table*}

The lowest-order coefficients $c_{0,1}$ do not vary drastically with $\beta$. 
We have found that the values are also similar to those obtained with just the two-parameter $k=0,1$ fits, 
i.e., they are rather robust, as expected. 
%
Furthermore, it is easily verified that, for the obtained parameters, we have the hierarchy 
\begin{equation} 
|\EPA_0| > |\EPA_1| > |\EPA_2| > |\EPA_3|
\end{equation} 
 within the density regime up to about $1$~fm$^{-3}$ for SNM \cite{npsm2017}
and up to about 0.05~fm$^{-3}$ for PNM, beyond which point we have $|\EPA_1|>|\EPA_0|$. 
Thus our physical reasoning holds up very well in SNM, while PNM deserves 
further investigation in the future.

Nonetheless, the dimensionless parameters which we defined in Eq.~(\ref{Eq:cdim}) do display naturalness. 
For SNM we obtain, for the ad-2 set:  
\[ 
c_0^{dim} = -3.6 , \quad c_1^{dim}=6.6, \quad c_2^{dim}=0.6  
\] 
and for PNM 
\[ 
c_0^{dim} = -1.1 , \quad c_1^{dim}=3.4, \quad c_2^{dim}=-5.9,  \quad c_3^{dim}=5.3 ~.
\]

\begin{figure*} 
\begin{center}
\includegraphics[scale=0.51]{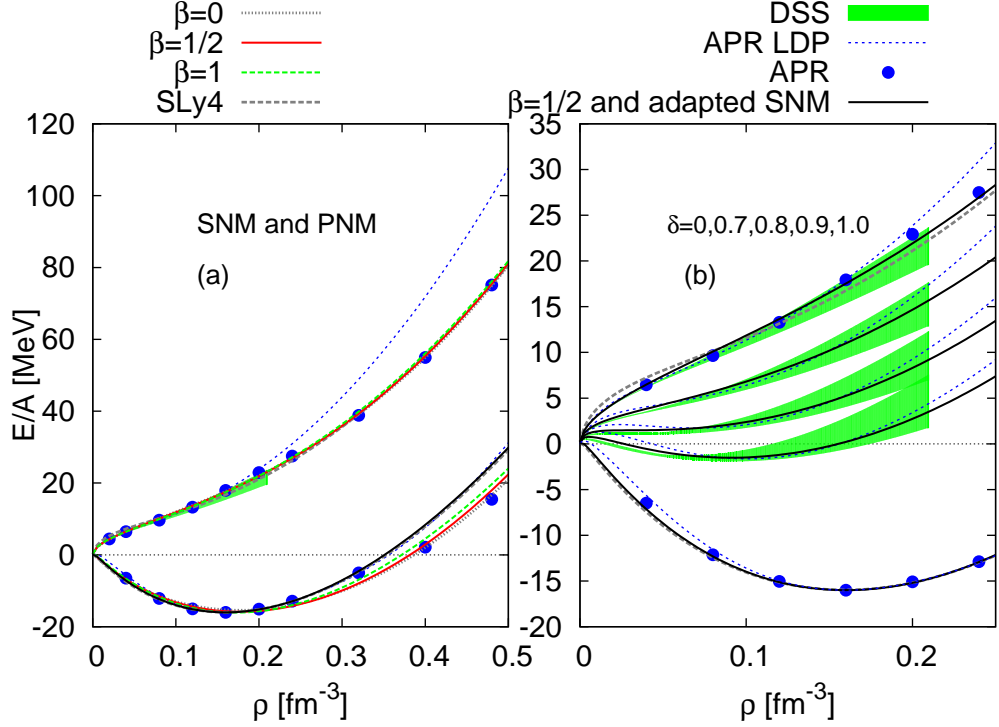}
\caption{\label{Fig:EoAb} 
Results of various fits are shown, for the energy per particle in SNM, PNM, and asymmetric matter, along with the APR pseudodata, the APR parameterization of the low-density phase (LDP), the DSS results from chiral EFT and the SLy4 functional. 
The fits of the terms $c_0,c_1,c_2,c_3$ with $\beta =0,1/2,1$ are included, as well as the parameterization with SNM parameters adapted to the saturation point, labelled "ad-1" in Table~\ref{Table:CI}. Results with the ``ad-2" set would be almost indistinguishable on the figure. 
%
%
}
\end{center}
\end{figure*} 

In Fig.~\ref{Fig:EoAb}, the results of various fits for the EoS are shown, along with the APR pseudodata, the APR parameterization of the low-density phase (LDP), the DSS results from chiral EFT and the SLy4 functional. 
In particular, 
Fig.~\ref{Fig:EoAb}(a) shows the energy per particle in SNM and PNM for the first four parameterizations from Table~\ref{Table:CI}. 
The last one, ad-2, is omitted because it gives almost indistinguishable results from ad-1. 

The discontinuity of the pseudodata around $\rho = 0.3$~fm$^{-3}$, which necessitated the weighted fits, is evident. 
The fitted functionals describe well not only the pseudodata, 
but also the DSS results, {\it which are not used in the fitting.} 
Fig.~\ref{Fig:EoAb}(b) shows clearly the excellent description of the DSS results for all four values of asymmetry.

\begin{figure} 
\begin{center}
\includegraphics[scale=0.40]{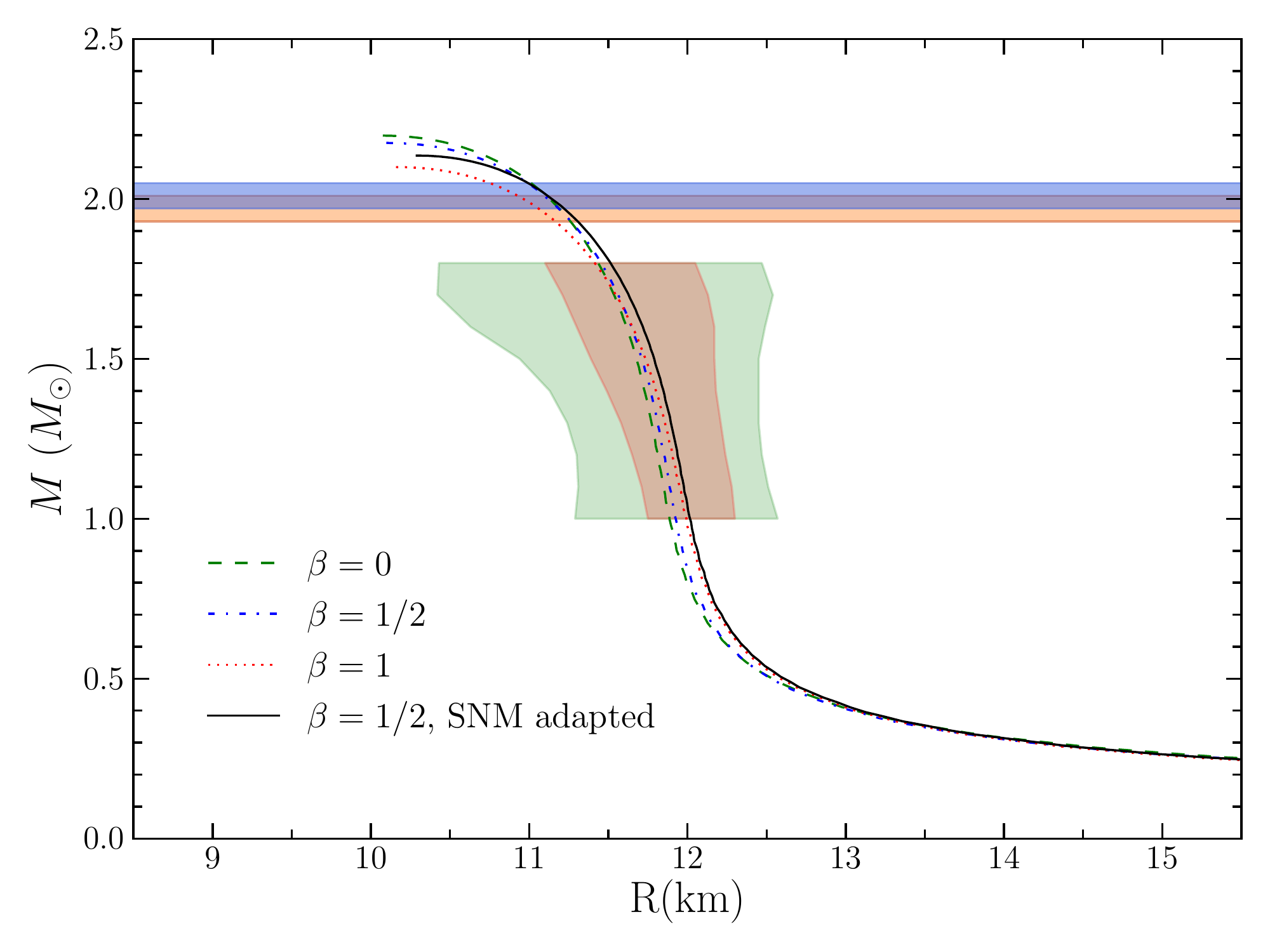}
\end{center}
\caption{\label{Fig:MR}Mass and radius relation of neutron stars for the models of Table~\ref{Table:CI}. 
The two horizontal bands represent the minimum of maximum mass
of neutron stars~\cite{Dem2010,Ant2013}. The central region
shows the allowed area of mass and radius of neutron stars
analyzed from X-ray burst data~\cite{SLB2010}.
The corresponding values for the central baryon density and the maximum mass $(\varrho_c,M)$ in units of (fm$^{-3},M_{\odot})$ are 
(1.135,2.20) ($\beta=0$), 
(1.140,2.18) ($\beta=1/2$), 
(1.165,2.10) ($\beta=1$), 
(1.135,2.14) ($\beta=1/2$, adapted to the saturation point of SNM). 
} 
\end{figure} 

The mass-radius relation for neutron stars obtained with the parameterizations of Table~\ref{Table:CI} is shown in Fig.~\ref{Fig:MR}. The corresponding values for the maximum mass and interior density is provided in the caption. All parameterizations predict values consistent with current constraints from observations~\cite{Dem2010,Ant2013}. 
We have verified that the SNM-adapted set along with the PNM parameters from the $\beta =1$ fit also produces
results within the desired constraints. 

We conclude that the power expansion of the nuclear EDF in Fermi momentum 
can provide an excellent description of symmetric and asymmetric nuclear matter in a large range of densities. 
In the future it is hoped that realistic, converged results will be derived within EFT, with which to compare our results{, or which will allow a thorough new study of PNM}.

\section{Summary and prospects\label{Sec:summary}}

We propose and explore a nuclear EDF written as a power expansion of the Fermi momentum. 
As such it is no less general than any available functional in analytical form. 
Although it can be viewed as an extended Skyrme EDF, the proposed form is not arbitrary and can be extended further to higher powers systematically.

Examining up to cubic terms, with the help of fits to microscopic calculations and a statistical analysis, 
we have verified the importance and robustness of low-order powers, {especially in SNM.} 
The resulting functional reproduces the known or adopted properties of saturated nuclear matter, dense matter, and neutron stars, as well as microscopic calculations for dilute matter, to which it was not fitted. 

The present work opens up different directions for further studies, some of which are currently underway: 

\begin{itemize} 
\item 
The proposed EDF can be recast in the form of a traditional Skyrme functional with gradient terms, for a variety of applications in finite nuclei. Our approach is to consider the obtained parameters $c_i$ already fixed in homogeneous matter and proceed to determine only the free parameters which cannot be constrained from unpolarized homogeneous matter. 
Minimally, these parameters are the {\it portion} of nonlocal {\it vs.} density-dependent terms (the sum being fixed for each power of $k_F$) and the spin-orbit force. The portion of momentum dependence in $c_2$ is related to the parameters $t_1,t_2$. 
They can both be determined, at least to a first approximation, from the ground-state energies and radii of closed-shell nuclei, as already done in Refs~\cite{Gil2016,npsm2017}.  
In the future, one can consider also pseudodata from  {\it ab initio} calculations of polarized matter. 
\item 
So far for our applications we have determined the parameterisations ad-1 and ad-2 with specific values for the SNM properties, in particular $K_{\infty}=240~$MeV. 
It will be interesting to vary this value to obtain a family of functionals. 
In the same spirit, one may also fix $J,L$, and so on without relying on pseudodata. One may also use different pseudodata sets. Such procedures will help us provide the uncertainties to all predictions. 
\item 
Combinations of higher-order terms in $k_F$ can also be explored in detail. 
Terms beyond $c_2$ may also be assumed to arise from momentum-dependent couplings in part.  
\end{itemize} 

We observed that the PNM shows different behavior from SNM: the hierarchy of terms is not as clear as for SNM. A convergence of the expansion for PNM is not evident from the present work. We find this observation worthy of future investigation in more fundamental ways.

\section*{Acknowledgments} 

The work of PP and YL was supported  by the Rare Isotope Science Project of the Institute for Basic Science funded by Ministry of Science, ICT and Future Planning and the 
National Research Foundation (NRF) of Korea (2013M7A1A1075764), 
of TSP by the Basic Science Research Program through the NRF
funded 
by the Ministry of Education, Science and Technology (NRF-2013R1A1A2063824) 
and 
of CHH by Basic Science Research Program through the NRF
funded by the Ministry of Education (NRF-2014R1A1A2054096).  

\section*{Appendix} 
For reference, in Table~\ref{Table:Chi2} we provide the $\chi^2$ results for all 31 combinations of 1,2,...,5 parameters up $k=3$ and including the logarithmic term.  There is overlap with Table~\ref{Table:Chi}. 
\begin{table}[h]
\begin{center}
\begin{tabular}{|r|c|c|c|c|}
\hline 
    &   $\beta=0$ & $\beta=\frac{1}{2}$ & $\beta =1$  &  $ \beta=\frac{3}{2}$  \\ 
\hline\hline 
    &     SNM   PNM   &   SNM   PNM   &  SNM   PNM  &  SNM  PNM   \\ 
\hline
         $k=0$  & 1.59534   0.39704  & 0.93074   0.17161  & 0.49065   0.07163  & 0.30167   0.03951  \\ 
         $k=1$  & 1.80178   0.34620  & 1.52783   0.22333  & 1.08948   0.13813  & 0.81692   0.10080  \\ 
         $k=2$  & 1.85877   0.29069  & 1.96985   0.24549  & 1.68528   0.19053  & 1.41039   0.15856  \\ 
         $k=3$  & 1.84307   0.24371  & 2.22864   0.24826  & 2.14357   0.22253  & 1.93646   0.20037  \\ 
(*)        $k=$ln  & 1.28821   0.41199  & 1.14335   0.18901  & 1.05040   0.13807  & 0.97133   0.12361  \\ 
\hline
       $k=0,1$  & 0.01304   0.02203  & 0.00387   0.00748  & 0.00115   0.00157  & 0.00073   0.00033  \\ 
       $k=0,2$  & 0.00936   0.00580  & 0.01227   0.00186  & 0.00944   0.00072  & 0.00526   0.00051  \\ 
       $k=0,3$  & 0.04116   0.00216  & 0.04777   0.00306  & 0.03583   0.00322  & 0.02199   0.00260  \\ 
       $k=1,2$  & 0.08530   0.00594  & 0.10870   0.00999  & 0.09030   0.01097  & 0.06219   0.00929  \\ 
       $k=1,3$  & 0.17598   0.01403  & 0.21642   0.02233  & 0.18341   0.02331  & 0.13480   0.01997  \\ 
       $k=2,3$  & 0.34238   0.03182  & 0.44056   0.04825  & 0.39801   0.05097  & 0.31871   0.04633  \\ 
(*)      $k=0,$ln  & 1.08517   0.24626  & 0.90154   0.16683  & 0.20457   0.03137  & 0.04192   0.00562  \\ 
(*)      $k=1,$ln  & 0.67040   0.09071  & 0.98501   0.13623  & 1.04965   0.13721  & 0.67757   0.06684  \\ 
(*)      $k=2,$ln  & 0.52929   0.05498  & 0.71715   0.08098  & 0.70648   0.08842  & 0.64413   0.08892  \\ 
      $k=3,$ln  & 0.50283   0.04942  & 0.66838   0.07285  & 0.62882   0.07735  & 0.53165   0.07304  \\ 
\hline
    $k=0,1,2$  & 0.00501   0.00329  & 0.00259   0.00178  & 0.00102   0.00053  & 0.00062   0.00014  \\ 
    $k=0,1,3$  & 0.00414   0.00192  & 0.00239   0.00119  & 0.00102   0.00040  & 0.00065   0.00012  \\ 
    $k=0,2,3$  & 0.00645   0.00206  & 0.00407   0.00154  & 0.00128   0.00064  & 0.00038   0.00022  \\ 
    $k=1,2,3$  & 0.02153   0.00518  & 0.01859   0.00516  & 0.00857   0.00302  & 0.00296   0.00153  \\ 
(*)   $k=0,1,$ln  & 0.00273   0.00032  & 0.00198   0.00026  & 0.00103   0.00014  & 0.00072   0.00007  \\ 
(*)   $k=0,2,$ln  & 0.00708   0.00155  & 0.00508   0.00137  & 0.00179   0.00068  & 0.00043   0.00028  \\ 
   $k=0,3,$ln  & 0.00912   0.00213  & 0.00668   0.00188  & 0.00238   0.00093  & 0.00053   0.00037  \\ 
(*)   $k=1,2,$ln  & 0.02690   0.00548  & 0.02606   0.00615  & 0.01422   0.00416  & 0.00597   0.00239  \\ 
   $k=1,3,$ln  & 0.03294   0.00664  & 0.03263   0.00754  & 0.01822   0.00510  & 0.00779   0.00288  \\ 
   $k=2,3,$ln  & 0.06537   0.01176  & 0.07466   0.01520  & 0.05067   0.01199  & 0.02660   0.00768  \\ 
\hline
   $k=0,1,2,3$  & 0.00162   0.00016  & 0.00173   0.00019  & 0.00101   0.00014  & 0.00037   0.00008  \\ 
(*)   $k=0,1,2,$ln  & 0.00147   0.00012  & 0.00163   0.00015  & 0.00102   0.00012  & 0.00040   0.00007  \\ 
   $k=0,1,3,$ln  & 0.00142   0.00012  & 0.00160   0.00014  & 0.00102   0.00012  & 0.00041   0.00007  \\ 
   $k=0,2,3,$ln  & 0.00111   0.00009  & 0.00127   0.00010  & 0.00087   0.00009  & 0.00038   0.00006  \\ 
   $k=1,2,3,$ln  & 0.00066   0.00019  & 0.00045   0.00014  & 0.00027   0.00010  & 0.00012   0.00008  \\ 
\hline
  $k=0,1,2,3,$ln & 0.00066   0.00009  & 0.00036   0.00008  & 0.00016   0.00006  & 0.00008   0.00005 \\  
\hline
\hline
\end{tabular} 
\end{center} 
\caption{\label{Table:Chi2}$\chi^2_n$ values for the indicated fits ($\beta=0,0.5,1,1.5$ and all combinations of non-zero $c_k$) of expression (\ref{Eq:FormC}) to the APR pseudodata in symmetric nuclear matter (SNM) and pure neutron matter (PNM). 
Sets marked with a star (*) contain the logarithmic term without a $k=3$ term, i.e., imply an arbitrary reference density $\rho_x=1$fm$^{-3}$.  
}
\end{table}

\end{document}